\documentclass{article}
\usepackage[utf8]{inputenc}
\usepackage{bbm}
\usepackage{breqn}
\usepackage{dsfont}
\usepackage{comment}
\usepackage{pgfplots}
\usepackage{authblk}
\usepackage{graphicx}
\usepackage{amsmath}
\usepackage{arydshln}
\usepackage[normalem]{ulem}
\usepackage[makeroom]{cancel}
\usepackage{tikz}

\usepackage{soul}
\pgfplotsset{compat=1.18}

\numberwithin{equation}{section}

\newcounter{example}
\newtheorem{example}{Example}[section]
\newcounter{remark}

\newcommand{\red}[1]{{\color{red} #1}}

\newcommand{\bra}[1]{\left\langle #1 \right|}
\newcommand{\y}[1]{\marginpar{\red{#1}}}
\newcommand{\braket}[2]{\langle #1 | #2 \rangle}
\newcommand{\ket}[1]{\left| #1 \right\rangle}
\title{Homotopy of periodic $2\times 2$ matrices }
\author[1,2]{Joseph E. Avron}
\author[1]{ Ari M. Turner}

\affil[1]{Department of Physics, Technion, Haifa, Israel}
\affil[2]{Helen Diller Quantum Center}
\date\today

\begin{document}

\maketitle
\begin{abstract}
We describe the homotopy classes of { loops in the space of} $2\times 2$ simple (=non-degenerate) matrices with various symmetries.  This turns out to be an elementary exercise in the homotopy of closed curves in $\mathbbm{R}^3/\{0\}$. Since closed curves in $\mathbbm{R}^3/\{0\}$ can be readily visualized, no advanced tools of algebraic topology are needed.  The matrices represent gapped Bloch Hamiltonians in 1D with a two dimensional Hilbert space per unit cell.
\end{abstract}



\section{Introduction}
We study the homotopy classes of loops in the space of simple $2\times 2$ matrices that satisfy various unitary or anti-unitary symmetries. This is motivated by the program of classifying the phases of insulating quantum matter \cite{haldane-nobel,kane-fu,kitaev,gu-wen}. The classification is enriched by allowing for  phases with symmetry \cite{zirn,rmp,schnyder,schnyder2}.

An insulator is characterized by a (many body) ground state separated {from the excited states} by a spectral gap which is bounded away from zero in the limit of large systems. The ground states of two insulators belong to the same phase if it is possible to interpolate between the two ground states without closing the gap. {One can picture a phase diagram in a space of Hamiltonians with all possible parameters; the problem of finding phases can be described topologically as finding the connected components of the region corresponding to Hamiltonians with a gap}\footnote{ Further topological properties of this space are also interesting; for example, its fundamental group plays a role in quantized pumping \cite{berg} and topologically protected quantum gates \cite{kitaev-anyons}.}.
This is a challenging mathematical program, especially for interacting systems, as it concerns many-body quantum mechanics in the thermodynamic limit  \cite{chen-gu-wen,levin,Tasaki,ogata,fid, pollmann-at,automorphic,Bachmann}.

The phases of topological insulators are of interest for the field of condensed matter because of their mobile edge states that are impervious to disorder \cite{halperin,graf} and
the search for quasi-particles with exotic statistics \cite{kitaev-anyons,oreg,nayak2008non}.  They are also of interest in quantum information theory because of their connections with computational complexity of many-body systems \cite{complexity}.

Arguably the simplest setting for studying phases of insulators is in the context of Bloch Hamiltonians\footnote{For the study of models of free Fermions in disordered systems see e.g. \cite{schulz, shapiro-thesis}.} for non-interacting 
Fermions \cite{haldane-nobel,kane-fu}. Because of their discrete translation symmetry they can be studied directly in the limit of infinite systems. {If the $m$ lowest bands (for some $m$) are separated from the rest by a gap, an insulator with $m$ Fermions per period can be formed by filling these states.} We focus on the simplest case where $m=1$ (with spinless Fermions) and where the dimension of space is one.

In one dimension, a Bloch band gives rise to a loop $P(k)$ of orthogonal projections where $k$ takes values in the Brillouin zone (BZ). The loop of band projections corresponds to a notion of a quantum state of a system with infinitely many particles. 

Suppose that one restricts oneself to Bloch bands with a given symmetry, e.g. time-reversal.   Two Bloch bands with projections $P_0(k)$ and $P_1(k)$ are said to be homotopic if there is a family of projections $P_t(k)$,  jointly continuous\footnote{The continuity follows from the  locality in real space of the Hamiltonian and the gap condition.} in $(t\in[0,1],k\in \text{BZ})$,  respecting the symmetry, that interpolates\footnote{An alternative definition of a homotopy uses the space of loops which has the topology known as the compact-open topology: a homotopy is a continuous path in this space.\cite{fox1945}} between $P_0(k)$ and $P_1(k)$. This corresponds to continuous changes between many-body quantum states of free Fermions.
Describing the equivalence classes of Bloch bands with various symmetries is an interesting and non-trivial theoretical problem.


Our modest aim  is to give an elementary invitation to this field by describing the homotopic classification of  non-degenerate two-band Bloch Hamiltonians in one dimension with various symmetries.
 The results, summarised in Table \ref{tab:periodic}, go somewhat beyond the first column in the periodic table of topological insulators\footnote{A comparison is made in Appendix \ref{app:Zirnbauer}.}\cite{kitaev,schnyder,schnyder2} for the Altland-Zirnbauer symmetry classes \cite{zirn} as we allow for certain space symmetries and some results are specific to two bands\footnote{See appendix \ref{app:Zirnbauer} for a comparison with the standard classification.}. 
Our methods are elementary, as we reduce the problem to the study of the homotopy of curves in three dimensions. The  elementary nature of the problem is a consequence of: 
\begin{itemize}
    \item The space of one dimensional   projections in $\mathbbm{C}^2$ is easy to visualize. It is simply the two sphere $\mathbbm{S}^2$. In contrast, the space of $n\times n$ rank $m$ projections,\footnote{This space is the Grassmannian, $Gr_m(\mathbbm{C}^n)$, a basic object in algebraic geometry and topology.} which describes insulators of $m$ occupied bands (see sec.~\ref{s:moredimensions}), is difficult to visualize.
    \item One-dimensional Bloch Hamiltonians with 2 gapped bands can be represented by closed curves in $\mathbbm{R}^3/\{0\}$ hence questions regarding the homotopy of Bloch Hamiltonians reduce to questions about the homotopy of curves in 3 dimensional Euclidean space with the origin removed. The study of the homotopy classes of curves in two and three dimensional spaces is the most intuitive part of homotopy theory. This is in contrast with the study of the homotopy of Bloch Hamiltonians in higher dimensions which translates to the study of the homotopy of higher dimensional closed surfaces which are not easily visualized and require more advanced tools\footnote{\samepage This is true even in the case of two gapped bands, since the high homotopy groups of the 2-sphere are non-trivial.}.
\end{itemize}

\begin{table}
    \centering
\begin{tabular}{ c| c | c}\label{t:pt}
 Symmetry & Constraint& Homotopy classes  \\ 
 \hline
 None & None &$\{\sigma_z\}$  \\  
$\Theta_+$ &$H(k)=H^*(-k)$& $ \{\sigma_z\}$ 
\\ 
$\Theta_-$ &$H(k)=\sigma_yH^*(-k)\sigma_y$& --- \\ 
$C_+$  &$H(k)=-{\sigma_z}H^*(-k){\sigma_z}$&$\left\{\pm {\sigma_x}, \ \pm  R_1\right\} $ 
\\
{$C_-$}  &$H(k)=-{\sigma_y}H^*(-k){\sigma_y}$&$ \{\sigma_z\} $  
\\ 
  B & $H(k)={\sigma_x}H(-k){\sigma_x}$ &$
    \{\pm {\sigma_x},  \pm R_1\}$ \\
S & $H(k)=G_kH(-k)G^*_k$ &$
   \{\pm \sigma_z\}
$ \\  \hdashline

$C_+\circ\Theta_+$ & $H(k)=-{\sigma_z}H(k){\sigma_z}$ &$\displaystyle{\{R_n|n\in\mathbbm{Z}\}}$
  \\
    $B\circ \,\Theta_+$ & $H(k)={\sigma_x}H^*(k){\sigma_x}$ & 
$\{R_n|n\in\mathbbm{Z}\}$\\
 $S\circ \,\Theta_+$ &$H(k)=G_kH^*(k)G^*_k$& $\{\pm \sigma_z\}$ 
\\
\hdashline
S $\land \,\Theta_+$ &$H(k)=G_kH(-k)G^*_k=H^*(-k)$&  $
    \{\pm \sigma_z\} 
$ \\  
B $\land \,\Theta_+$ &$H(k)={\sigma_x}H(-k){\sigma_x}=H^*(-k)$&  $\displaystyle{
  \{\pm R_n|n\in\mathbbm{Z}\}}  
$ 
\\    \hdashline
$C_+\land \Theta_+$& $H(k)=-{\sigma_z}H^*(-k){\sigma_z}=H^*(-k)$ &
  $\displaystyle{\{\pm R_n|n\in\mathbbm{Z}\}}$ 
  \\
  $C_- \land \Theta_+$& $H(k)=-{\sigma_y}H^*(-k){\sigma_y}=H^*(-k)$ &
  $\{\sigma_y\}$ 
  \\
\hfill\\
\end{tabular}

\caption{In the first column $\Theta_\pm$ stands for time-reversal with $\Theta_\pm^2=\pm 1$, similarly for Particle-Hole symmetry  $C_\pm$.  $B$ stands for bond-reflection and $S$ for site reflection.  $\land$ means that both symmetries are present and $\circ$ that only the product is a symmetry. $C_+\circ\Theta_+$ is chiral symmetry.  In the second column, ${\sigma_x},{\sigma_y}$  and ${\sigma_z}$ denote the Pauli matrices,  $*$ is complex conjugation,  and $G_k$ is the diagonal unitary of Eq.~\ref{e:G}.
The third column gives the equivalence classes $2\times 2$ periodic gapped matrices. $\{\sigma_x\},\{\sigma_y\},$ and $\{\sigma_z\}$ represent the trivial (= null-homotopic) Bloch matrix associated with a lattice of disconnected unit cells. The choice of Pauli matrix is not always canonical (see the relevant sections).  $R_n$ is the unitary given in Eq.~\ref{e:R}. 
The entries above the top dashed line correspond to generating symmetries.  The entries below show few composite symmetries.   The list of composites is incomplete: Only products with $\Theta_+$ are listed. }
    \label{tab:periodic}
\end{table}

\section{Bloch Hamiltonians with two bands}
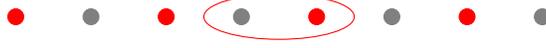
\begin{figure}[ht]\label{f:2uc}
    \centering
\begin{tikzpicture}{yscale=.6}

 \begin{scope}
\foreach \y in {2,4,6,8}
\filldraw [gray] (\y,0) circle (3pt);
\foreach \y in {1,3,5,7}
\filldraw [red] (\y,0) circle (3pt);
 \draw [red] (4+1/2,0) ellipse (1 and .3);  
 \end{scope}
\end{tikzpicture}
\caption{A periodic array with two atoms per unit cell.   The red ellipse shows a unit cell.} 
\end{figure}

Consider a one-dimensional, single particle Hamiltonian acting on the Hilbert space  
\begin{equation}
\ell^2(\mathbbm{Z})\otimes \mathbbm{C}^2
\end{equation}
This represents a chain of unit cells, each hosting a two dimensional Hilbert space, e.g. a spin or a pair of atoms; see Fig.~\ref{f:2uc}.  The basis states are $\ket{j,a}$ labeled by $j\in\mathbbm{Z}$ and $a\in
\{0,1\} $.

A one-particle lattice Hamiltonian is a self-adjoint  infinite dimensional matrix with elements
\begin{equation}
    \bra{j,b}H\ket{i,a}= \bra{i,a}H\ket{j,b}^*
\end{equation}
A Bloch Hamiltonian is, by definition,  translation invariant, i.e.
\begin{equation}
    \bra{j,b}H\ket{i,a}=\bra{0,b}H\ket{i-j,a}.
\end{equation}
Translation invariance implies that $H$ can be partially diagonalized by Fourier transform  (going to momentum space). For the Fourier transform to be well defined one needs to assume locality, i.e. 
\begin{equation}\label{e:summab}
   \sum_j\Big| \bra{0,b}H\ket{j,a}\Big|<\infty
\end{equation}
Locality, Eq. \ref{e:summab}, implies that the (reduced) Bloch Hamiltonian $H(k)$ defined by
\begin{equation}\label{e:FT}
    H_{ba}(k)=
    \sum_j \bra{j,b}H\ket{0,a} e^{-ikj}
\end{equation}
is a  continuous periodic matrix-valued function of $k$. $H(k)$ is Hermitian for each $k$ since
\begin{equation}\label{e:hermit}
       H_{ba}(k)=
    \sum_j \bra{j,b}H\ket{0,a} e^{-ikj}\\
    =H_{ab}^*(k)
\end{equation}
$H(k)$ is not uniquely defined: There is a $k$-dependent gauge freedom associated with the choice of unit cell as shown in the next section.
\subsection{Gauge ambiguity of (reduced) Bloch Hamiltonians}\label{s:unit-cell}
Changing the unit cell permutes the basis vectors, i.e.
\begin{equation}
    G^\ell\ket{j,b}=\ket{j- b \ell, b}, \quad b\in \{0,1\},\quad \ell\in \mathbbm{Z}
\end{equation}
See Fig.~\ref{f:2uc} for an illustration.  Suppose $\ell=1$.
\begin{figure}[ht]
    \centering
\begin{tikzpicture}[scale=2]
\begin{scope}
\foreach \y in {2,4,6,8,10}
\filldraw [gray] (\y/2,0) circle (1.5pt);
\foreach \y in {1,3,5,7,9}
\filldraw [red] (\y/2+1/2,1/2) circle (1.5pt);
 \draw [red] (2,1/4) ellipse (.3 and .7); 
 \begin{scope}[shift={(2.,2.85)}]
  \draw [red,rotate=-65] (3,1/4) ellipse (.3 and .8); 
 \end{scope}
 \end{scope}
\end{tikzpicture}
\caption{The two red ellipses correspond to two different choices of unit cell. Changing the unit cell is a gauge transformation (=diagonal unitary) acting on $H(k)$.}
\label{f:gauge}
\end{figure}
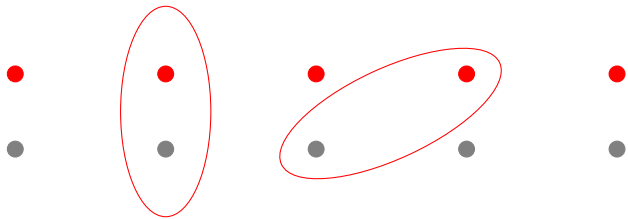
In coordinate space $H$ transforms by 
\begin{equation}
\bra{j,b}GHG^\dagger\ket{i,a}=\bra{j+b,b}H\ket{i+a,a}
\end{equation}
and $H(k)$ transforms by
\begin{equation}
    \big(G_k H(k)G_k^\dagger\big)_{b,a}=H_{ba}(k) e^{ik(b-a)}.
\end{equation}
The action of $G_k$ in $k$-space is implemented by a diagonal, k-dependent, unitary $2\times 2$ matrix:
\begin{equation}\label{e:G}
G_{k}= 
{\begin{pmatrix} 1 & 0 \\
  0&e^{ik}    \end{pmatrix}}
\end{equation}

\subsection{Periodic matrix valued functions and loops}\label{s:loops}

The space of Hermitian $2\times 2$ matrices is a linear space, with trivial topology (it is a contractible space).  In contrast, the space of simple (non-degenerate) $2\times 2$ matrices has a non-trivial topology. Since any simple matrix can be deformed to a traceless simple matrix, while maintaining simplicity, we assume, without loss, that the (reduced) Bloch matrices we consider are traceless\footnote{More precisely the space of matrices as a whole can be deformed by the homotopy
$f(H,t)=H-\frac{t}{2}(\mathrm{tr\ } H) \mathds{1}$ (where $t\in[0,1]$), which shows that the space of simple matrices is homotopy equivalent to the space of simple matrices with no trace.}. 

A traceless Hermitian $2\times 2$ matrix can be written as 
\begin{equation}\label{e:xyz}
    H=\mathbf{x}\cdot \boldsymbol{\sigma}
\end{equation}
where $\boldsymbol{\sigma}=(\sigma_x,\sigma_y,\sigma_z)$ is a vector of Pauli matrices.  The matrix is  simple (non-degenerate) if 
\begin{equation}
    \|\mathbf{x}\| >0.
\end{equation}
The space of simple $2\times 2$ (traceless) Hermitian matrices is Euclidean 3-space with the origin removed, which is  homotopic to the 2-sphere.

A gapped $2\times 2$ Bloch  Hamiltonian is then represented by a closed curve $\gamma$ in Euclidean 3-space with the origin removed:
\begin{equation}
\gamma=\{\mathbf{ x}_{k}| -\pi\le k \le \pi\} \in  \mathbbm{R}^3/\{0\}.
\end{equation}

Time-reversal, Particle-Hole, and reflection symmetries all relate $H(k)$ with $H(-k)$ and therefore $\mathbf{x}_k$ with $\mathbf{x}_{-k}$. In order to study deformation of $\gamma$ that respects the symmetry it is useful to think of the loop $\gamma$  as the concatenation of two curves
\begin{equation}\label{e:concatenate}
    \gamma=\gamma_-\circ\gamma_+
\end{equation}
where
\begin{equation}\label{e:half}
\gamma_-=\{\mathbf{ x}_{k}| -\pi\le k \le 0\}, \quad
\gamma_+=\{\mathbf{ x}_{k}| 0\le k \le \pi\}.
\end{equation}
with $\gamma_\pm $ related by the symmetry. 


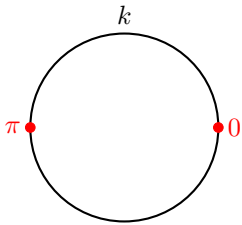
\begin{figure}[ht]
\centering
  \begin{tikzpicture}[scale=5/4]
  \draw [thick] (0,0) circle [radius=1] node [above] at (0,1) {$k$};
  \draw [fill, red] (1,0) circle [radius=.051] node [right] {$0$};
   \draw [fill, red] (-1,0) circle [radius=.051] node [left] {$\pi$};
\end{tikzpicture}
\caption{The one dimensional Brillouin zone is a circle parameterized by the angular  variable $k$. The time-reversal, Particle-Hole and reflection symmetry, relate $k$ with $-k$ and leave invariant the two points $k=0$ and $k=\pi$.}\label{f:flipk}
\end{figure}

\section{Symmetries}\label{s:prelim}
This section is a multilingual dictionary that translates several symmetries of $H$ in coordinate space to symmetries of $H(k)$ and then to symmetries of the loops $\gamma$. This will allow us  to determine the homotopy classes for each type of symmetry in the next section\footnote{Composed symmetries, such as chiral   symmetry, are relegated to Appendix \ref{a:cs}.}. (One can alternate between the two sections in order to understand each symmetry all at once.)


\subsection{Time reversal}

Time-reversal $\boldsymbol{\Theta}$ is an anti-unitary map. This is a consequence of  the presence of $i$ in the Heisenberg equation
\begin{equation}
   \dot{\mathbf{A}}= i[\mathbf{H},\mathbf{A}]
\end{equation}
where $\mathbf{A,H}$ are operators. Since 
$\{\boldsymbol{\Theta},\frac{d}{dt}\}=0$ 
it follows that $\{\boldsymbol{\Theta},i\}=0$ and so  $\boldsymbol{\Theta}$ is anti-linear.

Now for the problem studied in this article, where the operators act in Fock space, it is useful to represent time-reversal as an operation on single-particle states. By definition, time reversal maps particle creation to particle creation\footnote{The operation that maps creation to annihilation is Particle-Hole transformation.}, although a creation operator for a given state may map to a creation operator for some other state:
\begin{equation}   \boldsymbol{\Theta}a^\dagger(f)\boldsymbol{\Theta}^\dagger= a^\dagger\big(\Theta(f)\big).
\end{equation}
The bold and ordinary $\Theta$'s refer respectively to the transformation of the creation operator and the transformation of the wave function associated to it.
To derive the properties of the transformation $\Theta$ on the single particle wave functions from the properties of the many-body $\boldsymbol{\Theta}$, conjugate the anticommutation relation $\braket{g}{f}=\{a(g),a^\dagger(f)\}$ by $\boldsymbol{\Theta}$:
\begin{equation}
 \boldsymbol{\Theta}\braket{g}{f}\boldsymbol{\Theta}^\dagger=  \{\boldsymbol{\Theta}a(g)\boldsymbol{\Theta}^\dagger, \boldsymbol{\Theta}a^\dagger(f)
 \boldsymbol{\Theta}^\dagger\}=
    \{a({\Theta}g), a^\dagger({\Theta}f)\}=\braket{\Theta g}{\Theta f}
\end{equation}
Since $\boldsymbol\Theta$ is antiunitary, it acts as complex conjugation on a scalar, so the left-hand side is equal to $\langle f|g\rangle$ which forces $\Theta$ to be antiunitary.
Hence
\begin{equation}
    \Theta= U\circ *
\end{equation}
with $U$ unitary and $*$ complex conjugation.

{One reasonably assumes that} reversing time twice produces no
observable effect, i.e., it is a pure phase.  
Then it follows that, since $\Theta$ commutes with $\Theta^2$, the only phase consistent with the anti-unitarity of $\Theta$ is $\pm 1$, i.e.:
\begin{equation}
    \Theta^2=UU^*=\pm\mathds{1} \Longrightarrow U=\pm U^t
\end{equation}
 When $\Theta^2=\pm\mathds 1$, the symmetry will be denoted ${\Theta}_{\pm}$.
Under unitary change of bases $W$
\begin{equation}\label{e:cong}
    \Theta\mapsto W\Theta W^\dagger=WUW^t\circ *
\end{equation}
$U$ then undergoes a congruence transformation.

We assume that $U$ acts on the internal coordinates $\ket{a}$ {in real space while leaving the spatial coordinate $j$ unchanged.  It follows from $[H,\Theta]=0$ in real space that 
\begin{equation}
    \Theta H(k)\Theta^{-1}= UH^*(-k)U^\dagger
\end{equation}
Hence, $H(k)$ is time-reversal invariant if
\begin{equation}\label{e:anti}
    H(k)= UH^*(-k)U^\dagger
\end{equation}
for an appropriate unitary $U=\pm U^t$.
\subsubsection{$\Theta_+$}\label{s:trp}

For the symmetry $\Theta_+$, $U$ is a symmetric matrix.
Any symmetric unitary $2\times 2$ matrix $U$ is congruent to the identity\footnote{We thank Oded Kenneth for the proof below.}. This follows from the fact that $U$ can be written as
\begin{equation}
    U(\theta,\phi,\alpha)= e^{i\alpha} [\mathds{1} \cos\theta+ i\sin\theta ( {\sigma_x}\cos\phi+ {\sigma_z} \sin\phi)] 
\end{equation}
$U$ has a symmetric square root $\sqrt U= U(\theta/2,\phi,\alpha/2)$.
It follows that $U$ is congruent to the identity:
\begin{equation}
    U= \sqrt U\mathds{1} (\sqrt U)^t, \quad  
\end{equation}
Pick $U=\mathds{1}$, so
\begin{equation}\label{e:thetap}
    \Theta_+= *
\end{equation}
Then
\begin{equation}
    H(k)=H^*(-k)
\end{equation}
translated to a loop $\gamma$ with 
\begin{equation}\label{e:trs1}
     z_{k}= z_{-k}, \quad x_{k}= x_{-k}, \quad y_{k}= -y_{-k}
\end{equation}
$\gamma$ is symmetric under reflection in $y$
and is  anchored at $k=0,\pi$ to the punctured $x-z$ plane, see Fig.~\ref{f:gamma}.


\subsubsection{$\Theta_-$}\label{s:trn} 
There is a unique $2\times 2$ anti-symmetric unitary matrix $U$ up to multiplication by a scalar. Since an overall phase is for free in an anti-unitary transformation, we may chose $U={\sigma_y}$ 
 \begin{equation}\label{e:fermionic-tr}
    \Theta= {\sigma_y}\circ *, \quad {\sigma_y}=\begin{pmatrix}
     0&-i\\ i& 0
    \end{pmatrix}
\end{equation}
whose action on $H(k)$ is
\begin{equation}
    \Theta H(k)\Theta^{-1}=
    {\sigma_y} H^*(-k){\sigma_y}
\end{equation}
A $2\times 2$ matrix $H(k)$ is (Fermionic) time reversal symmetric if
\begin{equation}
    \mathbf{x}_k=-\mathbf{x}_{-k} 
\end{equation}
Since 
\begin{equation}
    \mathbf{x}_0=\mathbf{x}_\pi=0
\end{equation}
the Bloch Hamiltonian is gapless.
The family of Fermionic time reversal lies outside the gapped 2 bands framework and shall not be considered further.

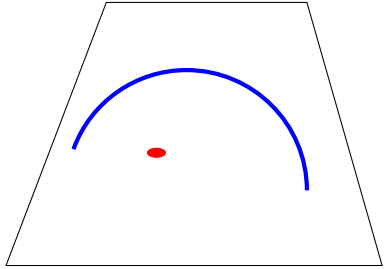
\begin{figure}
    \centering
    \begin{tikzpicture}[scale=2]
    \draw (-1,-1) -- (3/2,-1) -- (1, 3/4) -- (-1/3,3/4) --cycle;
    \draw[ultra thick, blue] (1,-1/2) arc (0:160:.8);
    \draw[fill,red] (0,-.25) circle (.06 and .03);
    \end{tikzpicture}
    \caption{The curve $\gamma_-$  associated with time-reversal symmetric Hamiltonians is anchored on the punctured $x-z$ plane. The rest of the loop is fixed by symmetry of reflection in the plane.}
    \label{f:gamma}
\end{figure}
\subsection{Particle-Hole transformation}

Particle-Hole transformation $\mathbf{C}$ maps particle creation to particle annihilation:
\begin{equation}
    \mathbf{C}a^\dagger(f)\mathbf{C}^\dagger= a\big(Cf\big)
\end{equation}
{The particle creation operator $a^\dagger(f)$ depends linearly on $f$ while the annihilation operator is anti-linear.  $\mathbf{C}$ (the many-body operator) must commute with $i$ in order to preserve the direction of time. Thus, in order for both sides to be linearly dependent on $f$, $C$ (the one-particle operator) must be anti-linear.
Being anti-unitary, $C$ comes in two varieties
\begin{equation}
    C^2=\pm\mathds{1}
\end{equation}
just like time-reversal.}
 ${C}_{\pm}$ denotes the symmetry corresponding to $C^2=\pm\mathds 1$.

\subsubsection{Particle-Hole symmetry}

Particle-Hole symmetry says that for any eigenstate of the single particle Hamiltonian $H$ there is a mirror state with an opposite energy and momentum. $C$  maps from the single particle state to its mirror image. {For Bloch Hamiltonians this means}
\begin{equation}\label{e:C}
{ (H)(k)+(CHC^{-1})(-k)=0}
\end{equation}
In contrast, the many-particle Hamiltonian $\mathbf{H}$ commutes with $\mathbf{C}$. To see this write
\begin{equation}
  \mathbf{H}=\sum \epsilon_{n,k}^{\phantom{\dagger}}a^\dagger_{n,k}a_{n,k}^{\phantom{\dagger}}   
\end{equation}
{ and then, formally denoting the mirror of band $n$ by $-n$,}
\begin{align}
    \mathbf{C}\mathbf{HC}^\dagger &=\sum \epsilon_{n,k}^{\phantom{\dagger}} 
    a_{-n,-k}^{\phantom{\dagger}}a_{-n,-k}^\dagger=\sum {\epsilon}_{-n,-k}^{\phantom{\dagger}} 
    a_{n,k}^{\phantom{\dagger}}a_{n,k}^\dagger  
    \nonumber
    \\
    &=-\sum \epsilon_{n,k}^{\phantom{\dagger}} 
    \big(\{a_{n,k}^{\phantom{\dagger}},a_{n,k}^\dagger\}-a^\dagger_{n,k}a_{n,k}^{\phantom{\dagger}}\big)= \mathbf{H} - \underbrace{\sum \epsilon_{n,k}}_{=0}\nonumber \\
    &=\mathbf{H}
\end{align}
It follows from Eq.~\ref{e:hermit}, the anti-unitarity of $C$ and Eq.~\ref{e:C} that a Bloch Hamiltonian has Particle-Hole symmetry if
\begin{equation}\label{e:c-sym}
    H(k)=-(CHC^{-1})({-}k)= -UH^*(-k)U^\dagger;\quad C=U\circ *.
\end{equation}
\subsubsection{$\mathbf{C}_+=1$}\label{s:cp}
By section \ref{s:trp} any anti-unitary transformation whose square is $\mathds{1}$ is equivalent to any another. Pick\footnote{Had we picked the same $U$ for time-reversal and Particle-Hole, the only Hamiltonian which is symmetric under both would have been $H=0$.} $U={\sigma_z}$, i.e.
\begin{equation}\label{e:ccp}
    C= \sigma_z\circ *
\end{equation}
The symmetry of $\gamma$ corresponding to 
\begin{equation}
    H(k)=-\sigma_z H^*(-k)\sigma_z
\end{equation} 
is
\begin{equation}\label{e:c1}
     z_{k}= -z_{-k}, \quad x_{k}= x_{-k}, \quad y_{k}= -y_{-k}
\end{equation}
$\gamma$ is symmetric under rotation by $\pi$ around the $x-$axis and is  anchored at $k=0,\pi$ to
the punctured $x-$axis, see Fig.~\ref{f:gamma-ct}.
\begin{figure}
    \centering
    \begin{tikzpicture}[scale=2]
    \draw (-1.2,-1) -- (1.7,-1) -- (1, 3/4) -- (-1/3,3/4) --cycle;
    \draw[ultra thick, blue] (1,-1/4) arc (0:180:.8);
    \draw[fill,red] (.25,-.25) circle (.06 and .03);
    \draw [->,thick] (-.7,-1/4)--(1.3,-1/4) node [below] {${x}$};
    \end{tikzpicture}
    \caption{The curve $\gamma_-$  associated with  Particle-Hole symmetric Hamiltonians with $C^2=\mathds{1}$ is anchored on the punctured $x$-axis. $\gamma_+$ is fixed by the symmetry as the $\pi$ rotation of $\gamma_-$ around the $x$-axis.}
    \label{f:gamma-ct}
\end{figure}
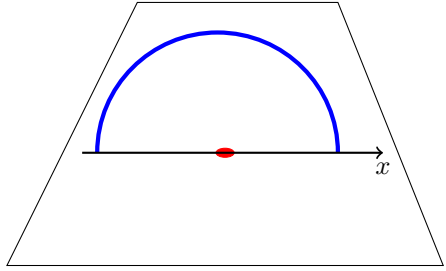
\subsubsection{$\mathbf{C}_-$} \label{s:ccn}
By section \ref{s:trn} we have:
\begin{equation}\label{e:flip-B}
    C= {\sigma_y}\circ * 
\end{equation}
whose action on $H(k)$ is
\begin{equation}
    C H(k)C^{-1}={\sigma_y} H^*(-k){\sigma_y}.
\end{equation}
So a $2\times 2$ matrix $H(k)$ is Particle-Hole symmetric if
\begin{equation}\label{e:phn}
    H(k)=-{\sigma_y}H^*(-k){\sigma_y}=H(-k)
\end{equation}
 The loop $\gamma$ corresponding to Eq.~\ref{e:phn} is self-retracing
\begin{equation}\label{e:gamma-fermionic-tr}
    \mathbf{x}_k=\mathbf{x}_{-k} 
\end{equation}
$ \mathbf{x}_0$ and  $\mathbf{x}_\pi$ can then be anywhere in $\mathbbm{R}^3/\{0\}$.

\subsection{Bond reflection symmetry}\label{s:brs}
\begin{figure}[ht]
    \centering
\begin{tikzpicture}{yscale=.6}
 \begin{scope}
\foreach \y in {2,4,6,8}
\filldraw [red] (\y,0) circle (3pt);
\foreach \y in {1,3,5,7}
\filldraw [red] (\y,0) circle (3pt);
 \draw [red] (4+1/2,0) ellipse (1 and .3); 
 \foreach \y in {1,3,5}
 \draw [thick,gray] (1.2+\y,0)--(1.8+\y,0);
 \end{scope}
 \draw [ultra thick,green,dashed] (4+1/2,-1)--(4+1/2,1);
\end{tikzpicture}
\caption{Bond reflection symmetry is a reflection about the green dashed line.}\label{f:brs}
\end{figure}
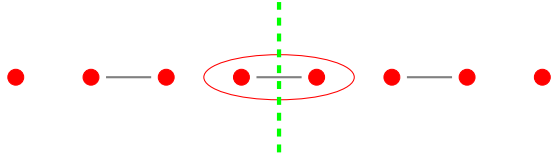
Bond reflection (see Fig.~\ref{f:brs}) permutes  the basis vectors
\begin{equation}\label{e:br}
    B\ket{j,a}=\ket{-j,a\oplus 1}, \quad B^\dagger=B
\end{equation} 
where $\oplus$ is addition modulo 2. The transformation reflects unit cells, and exchanges the ``atoms" in a unit cell. $B$ is an involution, $B^2=\mathds{1}$.

The action of bond-reflection on the Hamiltonian in the coordinate basis  is
\begin{equation}
    \bra{j,b}BHB^\dagger\ket{i,a}=\bra{-j,b\oplus 1}H\ket{-i,a\oplus 1}
\end{equation}
It follows from Eq.~\ref{e:FT} that the action in $k$-space is
\begin{equation}
    (BHB^\dagger)_{b,a}(k)= H_{b\oplus 1,a\oplus 1}(-k)
\end{equation}
In matrix form:
\begin{equation}\label{e:X}
    (BHB^\dagger)(k)=\begin{pmatrix} -z_{-k} & x_{-k}{+}iy_{-k} \\
  x_{-k}{-}iy_{-k}&z_{-k}    \end{pmatrix}={\sigma_x}H(-k){\sigma_x}
\end{equation}
$H(k)$ is bond-reflection symmetric if
\begin{equation}\label{e:brs}
    H(k)= {\sigma_x}H(-k){\sigma_x} 
\end{equation}
The loop 
$\gamma$ corresponding to Eq.~\ref{e:brs} is  {symmetric under} a rotation by  $\pi$ about the $x$-axis:
\begin{equation}
        z_k=-z_{-k}, \quad  x_k=x_{-k}, \quad  y_k=-y_{-k}
\end{equation}
and is anchored to the punctured $x-$axis, see Fig.~\ref{f:gamma-ct}.

\subsection{Site reflection symmetry}\label{s:srs}

\begin{figure}[ht]
    \centering
\begin{tikzpicture}{yscale=.6}

 \begin{scope}
\foreach \y in {2,4,6,8}
\filldraw [gray] (\y,0) circle (3pt);
\foreach \y in {1,3,5,7}
\filldraw [red] (\y,0) circle (3pt);
 \draw [red] (4+1/2,0) ellipse (1 and .3); 
 \end{scope}
 \draw [ultra thick, green,dashed] (4,-1)--(4,1);
\end{tikzpicture}
\caption{Site reflection symmetry is a reflection about the green dashed line}\label{f:srs}
\end{figure}
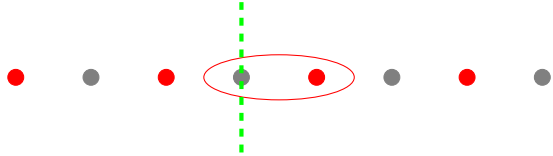
Reflection about a site (see Fig. \ref{f:srs}) permutes the basis vectors by
\begin{equation}\label{e:br1}
    S\ket{j,a}=\ket{-j-a,a}, \quad S=S^\dagger
\end{equation}
$S$ is an involution
that mixes different unit cells. Its action on $H$ in real space is given by
\begin{equation}
    \bra{j,b}SHS^\dagger\ket{i,a}=\bra{-j-b,b}H\ket{-i-a,a}
\end{equation}
By Eq.~\ref{e:FT}  site reflection acts in $k$-space by
\begin{equation}
    (SHS^\dagger)_{b,a}(k)= H_{b,a}(-k) e^{ik(b-a)}
\end{equation}
and in matrix form 
\begin{equation}\label{e:g}
    (SHS^\dagger)(k)=G_kH(-k)G^\dagger_k, 
\end{equation}
where $G_k$ is given in Eq.~\ref{e:G}.
Hence, $H(k)$ is site-reflection symmetric if
\begin{equation}\label{e:srs}
H(k)=G_kH(-k)G^\dagger_k 
\end{equation}
The symmetry of the loop corresponding to Eq.~\ref{e:srs} is a relation between 
 $\gamma_-$ and $\gamma_+$:
 \begin{equation}\label{e:com}
     z_k=z_{-k}, \quad \begin{pmatrix}
       x_k\\
       y_k
     \end{pmatrix}=  \begin{pmatrix}
       \cos k & -\sin k\\
       \sin k & \cos k
     \end{pmatrix} \begin{pmatrix}
       x_{-k}\\
       y_{-k}
     \end{pmatrix} 
 \end{equation}
 $\gamma$ is anchored at one end, $k=\pm\pi$, to the punctured $z$-axis
    \begin{equation}
      {  x_{\pm\pi}=y_{\pm\pi} =0, \quad |z_{\pm\pi}|>0}
    \end{equation}

\begin{example}
\label{ex:srs}
A  gapped Bloch Hamiltonian with site reflection symmetry is the SSH model \cite{ssh} with equal hopping and on-site potential:
\begin{equation}
    H(k)=\begin{pmatrix} v & 1+e^{-ik} \\
1+e^{ik} &-v    \end{pmatrix}, \quad v>0 
\end{equation}
\end{example}

\section{Homotopy of loops of gapped $2\times 2$ matrices  }\label{s:top}

Two Bloch Hamiltonians are homotopic if one can be deformed to the other, within a given symmetry class, without closing the gap. By section \ref{s:loops}  the question reduces to the homotopy of continuous closed loops  $\gamma\in\mathbbm{R}^3/\{0\}$ constrained by the symmetry. The symmetry constraint can sometimes complicate the question of homotopy equivalence, see section \ref{s:mirrorimages}. Fortunately, for most of the constraints we consider, this is not the case.  

{When the symmetries exchange $k$ and $-k$, the problem is simplified  by writing $\gamma$ as the concatenation $
    \gamma=\gamma_-\circ\gamma_+$ with $\gamma_\pm$ given by \ref{e:half}.
The symmetry says that $\gamma_-$ determines $\gamma_+$ (and vice versa), and imposes  constraints on the end-points\footnote{In some cases $\gamma$ is also constrained to be planar.} of $\gamma_\pm$.    Then any homotopy of the full curve $\gamma$ is determined by the homotopy of the arc $\gamma_-$, with the end-points required to satisfy the constraint. The homotopy equivalence of curves $\gamma_-$ is readily visualized and the determination of the equivalence classes reduces to an elementary exercise. Homotopy classifications for chiral symmetry are relegated to an appendix.    

\subsection{No symmetry}\label{s:ns}

$\mathbbm{R}^3/\{0\}$ is simply connected. This means that every (continuous) loop can be contracted to a point. Since $\mathbbm{R}^3/\{0\}$ is  connected, every point is homotopic to every other point.  Hence {any} loop of $2\times 2 $ gapped Bloch matrices $H(k)$ can be contracted to the constant $\sigma_z$ (for example), which represents a periodic array of isolated cells. 
 
The corresponding many-body ground state is a (formal) pure product state
\begin{equation}
    \prod_j \ket{0}_j
\end{equation}
{Being homotopic to a pure product states is taken as the defining property} of a trivial phase. This gives the first line in Table~\ref{t:pt}.

\subsection{ $\Theta_+$}

By section \ref{s:trp}  $\gamma$ is symmetric under reflection in $y-$axis, and is anchored to the $x-z$ plane for $k=0,\pi$.  Consider  the corresponding $\hat\gamma$ on the unit sphere whose end-points are anchored to the equator in the $x-z$ plane, see Fig.~\ref{f:gamma}.

$\hat\gamma_-$ can be deformed provided it respects the anchoring to the equator. $\hat\gamma_+$ follows by the symmetry constraint. Suppose first that the anchoring points coincide. Since $\mathbbm{S}^2$ is simply connected,  $\hat\gamma_-$ can be contracted to a point.  {The mirror image $\hat\gamma_+$ will contract simultaneously.}

In the case that the anchoring points do not coincide the { end-point at $k=\pi$ can be brought to the end-point at $k=0$ by applying a rotation around the $y-$axis $R_y(\phi)$ (since both lie on the unit circle in the $x-z$ plane). The angle $\phi$ is imagined to increase continuously from 0 {as a function of $k$ so that} the points coincide.} The $k$ dependent rotation
\begin{equation}
    R_y(k \phi/\pi)
    \hat{\mathbf{x}}_k, \quad 0\le k \le \pi
\end{equation}
is a homotopy of $\hat\gamma_-$ which respects the gap condition, since a rotation preserves the length of  vectors.
Hence $\hat\gamma_-$ is homotopic to a point.
Finally, since the equator in the $x-z$ plane is connected, all points on the equator are homotopic. 

It follows that the space of time reversal 1D Bloch Hamiltonians with a gap condition has a single component under homotopy, which we label by a single Pauli matrix $\{\sigma_z\}$. This accounts for the second line in table \ref{t:pt}.
\subsection{Particle-Hole symmetry}\label{s:charge}

\subsubsection{$C_+$} \label{s:php}
By section \ref{s:cp} the closed loop $\gamma$ is symmetric under rotation by $\pi$ about the $x-$axis, and is anchored to the (punctured) $x-$axis for $k=0,\pi$.
$\gamma$ is homotopic to $\hat\gamma$ on the unit sphere, anchored at the poles:
\begin{equation}
   \hat{\mathbf{x}}_0, \hat{\mathbf{x}}_\pi=  \pm(1,0,0)
\end{equation}
When the anchoring points coincide $\hat\gamma_-$ is  homotopic to a point.
In the case that the anchoring points are antipodal, 
$\hat\gamma_-$  can be continuously shortened to a geodesic. This means that $\hat\gamma$  is homotopic to a circle. 
\begin{figure}[ht]
\centering
  \begin{tikzpicture}[scale=6/5]
\draw [dashed] (1,1) -- (-1,-1) node {${\sigma_x}$};
  \draw [thick, blue] (0,0) circle [radius=1];
 \draw [cyan,thick,rotate=-0] (0,{sin(0)}) circle ({cos(0)}  and 1/4);
 \draw[ultra thick, blue] (.7,.7) arc (60:210:1.04);
  \draw[fill] (-.7,-.7) circle (.05 and .05);
  \draw[fill] (.7,.7) circle (.05 and .05);
 \draw [red,thick,] (.5,.5) circle (.3 and .2);
\end{tikzpicture}
    \caption{$\gamma_-$ with Particle-Hole symmetry. The blue spherical curve is anchored at $x=\pm 1$. The red one is anchored at $x=-1$. }
    \label{f:str}
\end{figure}
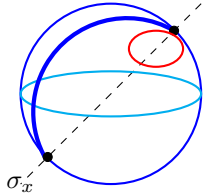
In conclusion, Particle-Hole symmetric Hamiltonians have four distinct equivalence classes represented by the four matrices:
\begin{equation}\label{e:R}
     \{ \pm {\sigma_x},\pm R_1\}, {\mathrm{where\ }}  
  R_n =\begin{pmatrix}
     0&e^{{-}ink}\\
     e^{ink}& 0
    \end{pmatrix}  
\end{equation}
($R\cong R^\dagger$ by a $\pi$ rotation about the $x$ axis.) {The four elements are distinguished by the combinations of signs of the anchoring points on $x$.}

$\pm {\sigma_x}$ is a Bloch Hamiltonian with disconnected cells, so the ground states are pure product states
\begin{equation}
    \prod_j \ket{+}_j,\quad  \prod_j \ket{-}_j, \qquad   \ket{\pm}= \frac{\ket 0\pm \ket 1}{\sqrt 2}
\end{equation}
representing trivial phases,
while the Hamiltonians $\pm R_1$ {when solved on a finite chain turn out to} have edge states at the ends.

\subsubsection{$C_-$}
By section \ref{s:ccn} the loop $\gamma$  is self-retracing and is not anchored.  Thus it can be contracted to a point along itself, so it is null-homotopic.  This gives the fifth line in table \ref{t:pt}.

\subsection{Bond reflection symmetry}\label{s:brs2}
By section \ref{s:brs} the curve $\gamma$ is anchored on the punctured $x$-axis. This is the same scenario as in  section \ref{s:php} and hence there are four equivalence classes labeled by the four matrices 
\begin{equation}
   \{  \pm {\sigma_x}, \ \pm R_1\}
\end{equation}
This gives the sixth line in table \ref{t:pt}.

\subsection{Site reflection symmetry}
\label{s:srsh}
By section \ref{s:srs} the loop $\gamma$ of Eq.~\ref{e:com} is anchored at one point, $k=\pm\pi$, to the punctured $z$-axis:
\begin{equation}\label{e:z-0}
 x_{\pm\pi}=y_{\pm\pi}=0, \quad |z_{\pm\pi}|>0
\end{equation}
Since $\gamma_-$ is anchored at the single point $k=-\pi$ it {can be contracted to the anchoring point by the homotopy
\begin{equation}\label{e:homotopy}
    \mathbf{x}_k(t)=\mathbf{x}_{ (1-t)k-t\pi}, \quad t\in[0,1]
\end{equation}
It follows that $\hat\gamma$ is homotopic to  either the north or south pole, and hence site symmetric Bloch Hamiltonians are homotopic to either one of two matrices 
\begin{equation}\label{e:z}
  \{ \pm {\sigma_z}\}
\end{equation}
The corresponding (many-body)  ground state represents a trivial phase as it is (formally) a pure product state.
This gives the seventh line in table \ref{t:pt}.
\begin{figure}[ht]
\centering
  \begin{tikzpicture}[scale=6/5]
  \draw [thick, blue] (0,0) circle [radius=1];
 \draw [cyan,thick,rotate=-0] (0,{sin(0)}) circle ({cos(0)}  and 1/4);
 \draw[ultra thick, blue] (0,1) arc (60:-30:1.2);
  \draw[ultra thick, red] (0,-1) arc (240:150:1.2);
  \draw[fill] (0,1) circle (.1 and .05);
   \draw[fill] (0,-1) circle (.1 and .05);
\end{tikzpicture}
\caption{Curves anchored at the north pole are homotopic to the north pole. And curves anchored at the south pole are homotopic to the south pole. This shows that the space of loops is made of two disconnected sets. }\label{f:n-s poles}
\end{figure}
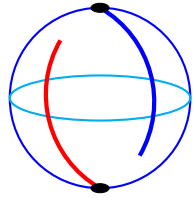

\subsection{$S\circ \Theta_+$:  The world-line of the loop.}\label{s:mirrorimages}

It is sometimes useful to consider the loop $\gamma$  as a world-line in the four dimensional space $\mathbbm{S}^1\times\mathbbm{R}^3$ where the first coordinate is $k$. A case in point is the composed symmetry $S\circ \Theta_+$, where the Bloch Hamiltonian satisfies the constraint
\begin{equation}
   H(k)=G_k H^*(k) G_k^\dagger ,\quad G_{k}= 
{\begin{pmatrix} 1 & 0 \\
  0&e^{ik}    \end{pmatrix}}
\end{equation}
This constrains the coordinates $(x_k,y_k)$:
\begin{equation}
    x_k-iy_k=(x_k+iy_k)e^{-ik} 
\end{equation}
andd gives no constraint on $z_k$. The constraint  implies
\begin{equation}
  (x_k,y_k)= \lambda_k \left( \cos \frac k2,\sin \frac k2\right),
\end{equation}
 where $\lambda_k$ is a real-valued function of $k$.

The loop $\gamma $ may be viewed as the world line in the  3-dimensional manifold    $\mathbbm{M}_{S\circ\Theta_+}$
\begin{equation}
 \mathbbm{M}_{S\circ\Theta_+}= \left\{\left(k,\lambda\cos\frac k2,\lambda\sin\frac k2,z\right )\Big |\lambda,z\in \mathbbm{R},-\pi\leq k\leq \pi\right\}
\end{equation}
embedded in the 4-dimensional space $\mathbbm{S}^1\times\mathbbm{R}^3$. $\gamma$ is not a closed curve in the $(\lambda,z)$ coordinates  because the $\lambda-z$ plane rotates relative to  the $x-y$ plane  by $\pi $  as $k$ goes from $-\pi $ to $\pi $.  The end-point $\mathbf{x}_{-\pi}=\mathbf{x}_\pi$ has different coordinates in the $(\lambda,z)$ plane
\begin{equation}
    \lambda_{-\pi}=-\lambda_{\pi}, \quad z_{-\pi}=z_\pi
\end{equation}

The homotopy classes of Hamiltonians
with $\mathrm{S}\circ\Theta$ symmetry
are thus homotopy classes of curves in a punctured $\lambda-z$ plane whose end-points are constrained to be
mirror-images in the $z$-axis, see Fig.~\ref{fig:atp}.
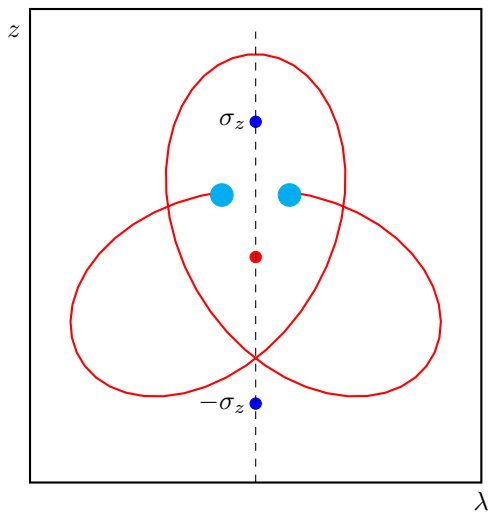
\begin{figure}
    \centering
   \begin{tikzpicture}[scale=3/2]
       \draw[thick] (-2,-2) rectangle (2,2.2);
       \draw [fill,red] (0,0) circle [radius=.05];
       \draw [fill,blue] (0,1.2) circle [radius=.05];
       \draw [fill,blue] (0,-1.3) circle [radius=.05];
       \draw [red, thick,  domain=-175:175, samples=80,scale=.6,rotate=450] 
 plot ({cos(\x) + 2* cos(2*\x)}, {sin(\x) - 2* sin(2*\x)} );
 \draw [dashed] (0,-2)--(0,2);
 \node[left] at (0,1.2) {${\sigma_z}$};
 \node[left] at (0,-1.3) {$-\sigma_z$};
 \node[below] at (2,-2) {$\lambda$};
  \node[left] at (-2,2) {$z$};
  \draw[fill, cyan] (-.3,.55) circle (.1);
   \draw[fill,cyan] (.3,.55) circle (.1);
   \end{tikzpicture}
    \caption{$\gamma$ in the $\lambda-z$ plane. The start and end points $(\lambda_{\pm\pi},z_{\pm \pi})$ are  the two cyan circles that are mirror imagees.  The red circle in the puncture in the plane.  The curve can be retracted to ${\sigma_z}$ by retracting the end points symmetrically relative to the $z$ axis. The curve can not be retracted to $-{\sigma_z}$.}
    \label{fig:atp}
\end{figure}
The end-points can be deformed
to be on the $+z$-axis (since any
point and its mirror
image can be moved continuously
to $+z$).  This curve is closed so
has a winding number. If one
of the end-points is taken around
the origin $n$ times
and brought
back (while
the other end-point mirrors the first), the
winding number
changes by $2n$. Thus, there are only two classes,
characterized by even and odd winding number
around the origin (after the end-points are moved
to the $+z$-axis).
The curves with even winding
number can be contracted
to constants $+\sigma_z$ and
the curves with odd winding number can
be contracted to $-\sigma_z$ (by moving the two end-points half-way around a circle in opposite directions)
so all Hamiltonians are homotopic to either $ \pm \sigma_z$\footnote{To distinguish between these two cases without first moving the end-points together, count the number of times the interior of the curve crosses the $-z$ axis. The parity is a homotopy invariant, and is equivalent to the previous classification.}.

\section{Changing unit cell permutes homotopy classes}
The labeling of the homotopy classes depends, in general, on the choice of unit cell \cite{kane-fu,ryu-hatsugai}. Different choices of unit cell are related by a gauge transformation, section \ref{s:unit-cell},
\begin{equation}
    H(k)\mapsto G_k^\ell H(k) G_k^{-\ell},  \quad \ell\in\mathbbm{Z}
\end{equation}
Under such a transformation the equivalence classes transform as
\begin{equation}
    {\sigma_z}\mapsto {\sigma_z}, \quad {\sigma_x}\mapsto R_\ell, \quad R_n\mapsto R_{n+\ell}. 
\end{equation}
This does not mean that the phases are equivalent, just that the labeling of the phase depends on how the unit cell is chosen. 

For example, in the case of bond-reflection symmetry,  the classes $\{\pm\sigma_x\}$ are exchanged with $\{\pm R_1\}$ under a gauge  transformation\footnote{The topological class of a perfectly periodic system has consequences for the behavior when the periodicity is broken by ending the chain, as long as far away from the edge the system is described by the periodic Hamiltonian. If the chain is stopped after a unit cell (whichever unit cell has been used to express the Bloch Hamiltonian), there will be an edge state for the $R_1$ {equivalence class} but not for the $\sigma_x$ {class}.}. This is the case in the Su-Schrieffer-Heeger model \cite{ssh}.  

\section{Fragile and New phases\label{s:moredimensions}}
The classification of Bloch Hamiltonians with two bands in 1D  can be extended to Bloch Hamiltonians with more bands. 
Fixing the number of bands to be $n$ and the number of  occupied bands to be $m < n$, we are interested in the homotopy of loops of $n\times n$ invertible Hermitian matrices with $m$ strictly
negative and $n-m$ strictly positive eigenvalues.

The homotopic classification in  the general case may differ from the classification of two bands: First, distinct phases in the two bands case may dissolve into one equivalence class upon the addition of bands. 
Second, new phases may appear.

\subsection{A fragile phase}\label{sec:fragile}
{A phase (with a given symmetry)} is fragile \cite{fragile} if its homotopy class changes when the dimension of the  matrices of band projection $P(k)$ is increased\footnote{Usually, as the dimensions of the Hamiltonians are increased, the classification eventually stops changing, and are called stable classes in K-theory. The fragile phases are the unstable phases that exist only in sufficiently small dimensions.  }:  
\begin{equation}
    P(k)\mapsto \begin{pmatrix}
        P(k)& 0\\
        0& 0
    \end{pmatrix}
\end{equation}
As an example, of a fragile topological phase consider an $n\times n$ real, periodic,  projection  $P(k)$ onto a single band: 
\begin{equation}
    P(k)=P^*(k)= P(k+2\pi)=P^2(k), \quad Tr\,P(k)=1
\end{equation}
The  reality condition may be viewed to be a consequence of the symmetry constraint $B\circ\Theta_+$ upon choosing $\Theta_+=\sigma_x\circ *$.

Since $P(k)$ is a real symmetric matrix it can be diagonalized by an orthogonal transformation and its non-trivial eigenvector
\begin{equation}
    P(k)\ket{k}=\ket{k}
\end{equation}
can be chosen to be real vectors of unit length in $\mathbbm{R}^n$. The eigenvalue equation then determines $\ket k$ uniquely, up to a $\pm 1$ phase ambiguity. Since $P(k)$ is continuous  in $k$, we may choose $\ket k$ so that it is a continuous family for $k\in [-\pi,\pi]$.  Then $\ket k$ traces a continuous curve on the unit sphere in $\mathbbm{R}^n$. Since $P(-\pi)=P(\pi)$ and $\ket k$ has $\pm 1$ phase freedom, the curve is either a closed loop or one with antipodal end-points:
\begin{equation}\label{e:sa}
    \ket{k=-\pi}=\pm \ket{k=\pi}
\end{equation}
The $\pm$ alternative divides the space of curves $\ket k$ on the sphere into two classes: A symmetric class where $\ket{k=\pi}=\ket{k=-\pi}$ and an anti-symmetric class where $\ket{k=\pi}=-\ket{k=-\pi}$. This holds for any $n\ge 2$. 

{In the case $n=2$ the curves can be classified further according to the winding number of $\ket k$ on the circle, which is a whole number (for the symmetric class) or a half-integer for the anti-symmetric class, see Fig.~\ref{f:vectors}. }
This can be related to the winding number of the curve $\gamma$ representing the Hamiltonian: Since $H$ is real, $\gamma$ is in the $x-z$ plane. $\gamma$ winds twice as fast as the eigenvector, so its winding number is an integer (since the Hamiltonian is periodic).  It is odd for an anti-symmetric eigenvector and even for a symmetric one. 

{For $n\ge 3$ the vector $\ket k$ lies on $\mathbbm{S}^{n-1}$
which is simply connected, so all {integer} winding can be undone, reducing to a single class for the symmetric case and a single class for the anti-symmetric case. }

{Thus, we find that the periodic real projections $P(k)$ are labeled by:
\begin{equation}
  \mathbbm{Z} \text{ for } n=2, \quad  \mathbbm{Z}_2 \text { for }  n\geq 3   
\end{equation}
 {Hence all real Bloch bands with an even winding number of $\gamma$ collapse to a single equivalence class, 
 while those with an odd winding number collapse to a second equivalence class,} once the dimension of the Hamiltonian is increased above 2.}

\begin{figure}[ht]
\centering
  \begin{tikzpicture}[scale=5/4]
  \begin{scope}
  \draw [fill, red] (0,0) circle [radius=.051];
   \draw [fill, red] (-2,0) circle [radius=.051];
   \begin{scope}
        \draw[thick, blue] (0,0) arc (0:180:1);
        \draw[thick, dashed,blue] (0,0) arc (0:-180:1);
   \end{scope}
    \begin{scope}[scale=.95,shift={(-.05,0)}]
        \draw[thick, black] (0,0) arc (0:360:1);
   \end{scope}         
  \end{scope}
  \begin{scope}[shift={(4,0)}]
  \draw [fill, red] (0,0) circle [radius=.051];
   \draw [fill, red] (-2,0) circle [radius=.051];
   \begin{scope}
        \draw[thick, blue] (0,0) arc (0:180:1);
        \draw[thick, dashed,blue] (0,0) arc (0:-180:1);
   \end{scope}
    \begin{scope}[scale=.95,shift={(-.05,0)}]
        \draw[thick, black] (0,0) arc (0:360:1);
   \end{scope}  
    \draw[thick, black] (-1,0) circle (1 and 1/4);
  \end{scope}

\end{tikzpicture}
\caption{The blue solid semi-circle and the dashed semi-circle are homotopically distinct when the embedding space is the plane but are  equivalent in three dimensions. }\label{f:vectors}
\end{figure}
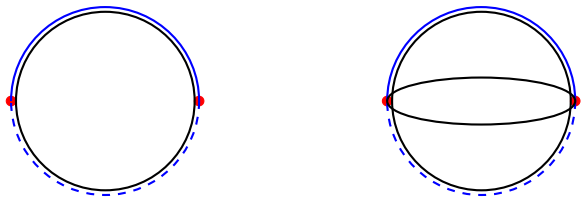

\subsection{New phases\label{sec:new}}

As an example of new phases that emerge when the number of bands is increased consider the reflection  symmetry  ${R}$.  The points $k=0,\pi$ are special as
\begin{equation}
    [{R},H(k)]=0, \quad k\in\{0,\pi\} 
\end{equation}
Hence, eigenstates of $H(k),$  for $k=0,\pi $ are also eigenvectors of ${R}$.
Define  the index $N_-(k)\in\{0,\dots,m\}$ as the sum of negative states with negative parity:
\begin{equation}
    N_-(k)=\sum_{ \epsilon_n<0} \frac{1-\bra{\psi_n}{R}\ket{\psi_n}} 2, \quad H(k)\ket{\psi_n}=\epsilon_n \ket{\psi_n}, \quad k\in\{0,\pi\}
\end{equation}
$N_-$ is an invariant under deformations of $H(k)$ that respect the parity and do not close the gap\footnote{The number of negative energy states with eigenvalues $+1$ is not a separate invariant since the total number of negative eigenvalues is fixed, see \cite{Inversion-paper}.}.

In the case of $m$ occupied bands one expects to find, in general, $(m+1)^2$ equivalence classes labeled by $(N_-(0),N_-(\pi))$ and the number of invariants to increase with the number of the bands.

It is instructive to examine the results in Table \ref{t:pt} where $m=1$, from this perspective:
\begin{itemize}
    \item In the case of bond reflection $R_k=\sigma_x$ and 
    $N_-(k)\in \{0,1\}$. Table \ref{t:pt} indeed gives four equivalence classes, as it should. 
\item In the case of site reflection $R_k=G_k$  is the identity for $k=0$  so automatically $N_-(0)=0$ and the two equivalence classes are labelled by $N_-(\pi)\in \{0,1\}$  in agreement with Table~\ref{t:pt}.
\end{itemize}

\subsection{Stable phases}

{\color{black}Since the number of invariants may either shrink or grow when the number of bands increases, it is natural to introduce a notion of ``stable invariants". For sufficiently large (invertible) matrices where both the numbers of positive and negative eigenvalues are large, the topological classification often stops changing, and the ``stable invariants" describe these classes.  We shall not discuss further this issue. }

\section{Concluding remarks}

Our aim here was to give an elementary invitation to the field of topological insulators by studying the homotopic classification of $2\times 2$ simple (=non-degenerate) periodic matrices in one dimension with various symmetries.  Using simple intuitive facts about the homotopy of curves, we have shown how various symmetries are expressed in the homotopic classification of Bloch Hamiltonians. We have also presented an introduction to the notions of fragile and stable topological phases.

\section*{Acknowledgement} We thank  Jacob Shapiro for careful reading of an earlier version of the paper and many helpful suggestions.  JA also thanks Johannes Kellendonk, Oded Kenneth, Roy Meshulam, Daniel Podolsky, Immanuel Bloch and Raffaele Resta for useful discussions.  AMT acknowledges the ISF for support under the grant ISF 1939/18.
\appendix

\section{Various symmetry compositions}
\subsection{Chiral symmetry:   $C_+\circ\Theta_+$}\label{a:cs}
Chiral symmetry (aka sublattice symmetry) is the \emph {composition} of Particle-Hole and time reversal symmetry. As such it is a unitary map that preserves $k$,
\begin{equation}
    H(k)\mapsto - U H(k)U^\dagger, \quad U=U_c U_t^* 
\end{equation}
where $\Theta_+=U_t\circ *$ and $C_+=U_c\circ *$.  {\color{black} Note that, in order for chiral symmetry to lead to an interesting classification, one must choose $U_t\neq U_c$.   If $U_t=U_c$, this would have the unfortunate consequence that $U=U_cU^*_c=U_c U^\dagger_c=\mathds{1}$ and chiral symmetry would imply $H(k)=0$.} 

Choosing $\Theta_+$ and $C_+$ as in Eqs. \ref{e:thetap},\ref{e:ccp}, $H(k)$ has chiral symmetry if
\begin{equation}
    H(k)=-  \sigma_zH(k)\sigma_z
\end{equation}
This says that the curve $\gamma$ lies in a punctured $x-y$ plane\footnote{Had we chosen $U=\sigma_y$ the curve $\gamma$ would lie in the $x-z$ plane and $H(k)$ would then be a real matrix.}, which is not simply connected and has its winding number as a topological invariant
\begin{equation}
    \text{Winding}(\gamma)\in \mathbbm{Z}
\end{equation}

$H(k)$ with winding $n$ is homotopic to $R_n$ of Eq.~\ref{e:R}. It follows that chiral Bloch Hamiltonians are homotopic to
\begin{equation}
 \{R_n|n\in\mathbbm{Z}\}
\end{equation}

\subsection{$C_+{\color{black}\land}\Theta_+$}}


Particle-Hole and chiral symmetry together also imply symmetry under their composition, chiral symmetry.  Chiral symmetry restricts $\gamma$ to the  $x-y$ plane and time reversal forces the anchoring points to lie in the $x-z$ plane.  It follows that the anchoring points at $0,\pi$ must lie on a punctured $x$-axis. {\color{black} Although $\gamma$ and  $-\gamma$ have the same winding numbers and can be rotated into one another by applying a $\pi$ rotation about the $z$-axis, they are  homotopically distinct since such a rotation does not respect the anchoring condition}.  This doubles the homotopy classes
  \begin{equation}
R_n\mapsto \pm R_n. 
\end{equation}
 To summarize, the homotopy equivalence classes are given by
\begin{equation}
    \{\pm R_n|n\in\mathbbm{Z}\}
\end{equation}
\subsection{$B\circ \Theta_+$} For Hamiltonians with only one symmetry, the \emph{composition} of time reversal and bond reflection, one has
\begin{equation}\label{e:bt}
    H(k)= (B\Theta H \Theta^{-1}B)(k)=\sigma_xH^*(k)\sigma_x
\end{equation}
which implies
\begin{equation}\label{e:zk}
    z_k=0
\end{equation}
so  $\gamma$ lies in the punctured $x-y$ plane, and there is no constraint on the anchoring points at $k=0,\pi$.  The equivalence classes are labelled by their winding number and all Bloch Hamiltonians with winding $n$ are homotopic to the matrix $R_n$. This gives the ninth line in table \ref{t:pt}: 
\begin{equation}
\{ R_n|n\in\mathbbm{Z}\}.
\end{equation}
\subsection{$B\land \Theta_+$}

Now, in addition to Eqs.~\ref{e:bt},\ref{e:zk} bond reflection gives the additional constraint that $\gamma$ is anchored on the punctured $x-$axis.  It follows that $\hat\gamma$ is  distinguished  by its winding, and the anchoring point $\mathbf{x}_0=(\pm 1,0,0)$.  The equivalence classes are represented by the matrices
\begin{equation}
\{\pm R_n|n\in\mathbbm{Z}\}
\end{equation}
This gives the twelfth line in the table \ref{t:pt}.

\subsection{$S\land \Theta_+$\label{sec:srtr}}

When $S$ and $\Theta_+$ symmetries are combined, then (as in section \ref{s:mirrorimages}) the curve $\gamma$ is not in a fixed space, independent of $k$.
This is not anymore the case for 
Bloch Hamiltonians with time reversal and site reflection symmetry. 

The conditions are that
\begin{equation}\label{e:constraint}
  (x_k,y_k)= \lambda_k \left( \cos \frac k2,\sin \frac k2\right),\quad (\lambda_k,z_k)=(\lambda_{-k},z_{-k}), \quad {\lambda_{\pm\pi}=0}
\end{equation}
The first condition follows from the site-reflection condition { Eq.~\ref{e:com}}, as in section \ref{s:mirrorimages}. Time-reversal implies that $x_k=x_{-k}$ and
$y_k=-y_{-k}$; i.e., $\lambda_k=\lambda_{-k}$. Also,
$\lambda_{\pm\pi}=0$ by Eq.~\ref{e:z-0}.
$\gamma_-$ is anchored at {$-\pi$} on the punctured $z$-axis.  The arc can be contracted to the point it is anchored to on the north or to the south pole while respecting the constraint, by contracting it along itself in the $z-\lambda$ space.  I.e. for $k<0$,
\begin{equation}
    z_k(t)= z_{(1-t)k -t \pi} , \quad
     (x_k,y_k)= \lambda_{(1-t)k-t\pi} \left( \cos \frac k2,\sin \frac k2\right)
\end{equation}
There are two homotopy classes determined by the anchoring points: 
\begin{equation}
 \{\pm {\sigma_z}\}
\end{equation}
This gives the fourth line from the bottom of table \ref{t:pt}.

\section{Comparison with the standard classification\label{app:Zirnbauer}}

The standard classification of topological insulators is concerned with the 10 types of symmetry that do not have spatial symmetries other than translations. It is made from all the possible combinations of time reversal and Particle-Hole symmetry. The standard classification and the two-band classification are compared in the table below.

\begin{itemize}
\item The standard classification labels the homotopy equivalence classes by Abelian groups where $0$ says that there is one equivalence class.  
Since Bloch Hamiltonians do not come with a natural group structure, we found it more intuitive to mark the equivalence classes by a particularly simple Bloch matrix in the set. This  choice is constrained by the symmetry, but otherwise it is arbitrary. For example, in the first line of the table, any Pauli matrix could be chosen (for example), but in the third row the $\Theta_+$ symmetry allows only real matrices. The number of elements in the third column gives the number of homotopically distinct sets.

    \item In the two cases marked by $*$ in first column the two band classification gives twice as many equivalence classes as in the standard classification.  The extra  $\mathbbm{Z}_2$  can be traced to the sign of the Pfaffian  of $-iH$ where $H$ is Particle-Hole symmetric matrix (and so imaginary and anti-symmetric).  The standard classification disregards this ``weak" $\mathbbm{Z}_2$ index because it is a zero-dimensional property.
    \item 
In three cases, marked by $-$ in the second column, no comparison is made as there are no gapped 2-band Hamiltonian with $\Theta_-$ symmetry. 
\end{itemize}
 For these symmetries, there are no fragile phases, which would have also caused there to be more phases in our classification than in the standard classification. Fragile phases do occur when there is reflection symmetry, see Sec. \ref{sec:fragile}.
 
\begin{table}
\centering
\begin{tabular}{c|c|c|c}
{\phantom{x}}& {Symmetry}& {Two-Band Classification} & Standard Classification\\
\hline
A& None&0&0 \\
AIII &Chiral&$\mathbbm{Z}$&$\mathbbm{Z}$\\
AI&$\Theta_+$&0&0\\
BDI$^*$&$\Theta_+,$C$_+$&$\mathbbm{Z}_2\times\mathbbm{Z}$&$\mathbbm{Z}$\\
D$^*$& C$_+$&$\mathbbm{Z}_2\times\mathbbm{Z}_2$&$\mathbbm{Z}_2$\\
DIII$^\dagger$ &$\Theta_-$,C$_+$&--&$\mathbbm{Z}_2$\\
AII$^\dagger$&$\Theta_-$&--&0\\
CII$^\dagger$&$\Theta_-$,C$_-$&--&$\mathbbm{Z}$\\
C&C$_-$&0&0\\
CI&$\Theta_+$,C$_-$&0&0\\
\hline
\end{tabular}

\caption{Comparison between standard classification of topological insulator classification in one dimension and classification for $2\times 2$ Hamiltonians, see Table \ref{t:pt}. The classifications agree except that in the standard classification a weak $\mathbbm{Z}_2$ index is ignored in two cases (marked $^*$), corresponding to the choice of $\pm$ in the Hamiltonians, and whenever there is $\Theta_-$ symmetry (marked $^\dagger$), there are no $2\times 2$ Hamiltonians at all that are gapped.}
\end{table}


\end{document}